\documentclass[12pt,fleqn,a4paper]{article}
\usepackage{pstricks}
\usepackage{pst-node}
\newcommand{\beq}{\begin{eqnarray} \displaystyle}
\newcommand{\eeq}{\end{eqnarray}}
\newcommand{\summe}[1]{\sum\limits_{#1}}
\newcommand{\sumijs}{ \sum\limits_{i\neq j\sigma}}
\newcommand{\cplus}[1]{{\bf c}^+_{#1}}
\newcommand{\cminus}[1]{{\bf c}_{#1}}
\newcommand{\cpsigma}[1]{{\bf c}^+_{#1 \sigma}}
\newcommand{\csigma}[1]{{\bf c}_{#1 \sigma}}
\newcommand{\nup}[1]{{\bf n}_{#1 \uparrow}}
\newcommand{\ndown}[1]{{\bf n}_{#1 \downarrow}}
\newcommand{\nsigma}[1]{{\bf n}_{#1 \sigma}}
\newcommand{\noperator}[1]{{\bf n}_{#1 }}

\newcommand{\hamilton}{\mbox{\boldmath $H$}}

\newcommand{\liouville}{\mbox{\boldmath $L$}}

\newcommand{\operator}[1]{\mbox{\boldmath $#1$}}

\newcommand{\opA}{\mbox{\boldmath $ A $}}
\newcommand{\opB}{\mbox{\boldmath $ B $}}
\newcommand{\opC}{\mbox{\boldmath $ C $}}
\newcommand{\opD}{\mbox{\boldmath $ D $}}
\newcommand{\opL}{\mbox{\boldmath $ L $}}

\newcommand{\mket}[1]{| #1 \rangle}
\newcommand{\mbra}[1]{\langle #1|}
\newcommand{\ket}[1]{| #1 \rangle}
\newcommand{\bra}[1]{\langle #1|}
\newcommand{\expect}[1]{\langle #1 \rangle}
\newcommand{\ketvacuum}{\mket{\mbox{vac}}}
\newcommand{\bravacuum}{\mbra{\mbox{vac}}}

\newcommand{\expo}[1]{e^{\displaystyle #1}}

\begin{document}
\begin{center}
{\Large\bf 
Perturbation Expansion of the Partition Sum 
for any Temperature \\[2ex]} 
R. Schumann \\[1ex]
{\small Institut f\"ur Theoretische Physik, TU Dresden, 
D-01062 Dresden, Germany}\\
{\small e-mail: schumann@theory.phy.tu-dresden.de } \\[1.0cm]
\begin{minipage}[t]{13cm}
\small
\begin{center} 
\bf Abstract
\end{center}
Based on the special properties of Liouville eigenoperators a perturbation
theory for the partition sum is given. It is applicable for any 
temperature and includes the case of degenerate Hamiltonians. 
To demonstrate the 
realibility of the method, the second order correction to the atomic limit 
grand canonical potential of the Hubbard model is calculated and compared
to results known from the literature.
\end{minipage}\\[1cm]
\end{center}
\section{Introduction}
From the early beginning of statistical thermodynamics 
there was no doubt that 
the partition sum is the central
quantity for describing the equilibrium quantities in physics.
In spite of its central role there is no easy to handle perturbation theory
for arbitrary temperature up to now. Of course, from the first days of
quantum mechanics there were attempts in this direction. They all suffer 
from the non-commutativity of the perturbation with the unperturbed 
Hamiltonian,
which makes it necessary to introduce an imaginary time via the Feynmann
time ordering trick to split the exponential contained in the partition sum,
resulting in various graph schemes for perturbation series or two time
Green functions. Of course, perturbation theories are widely employed
in all fields of contemporary physics and, therefore, it is hopeless to review
all the developments, so I restrict to a brief review of the state of art in
the context of the Hubbard model \cite{Hubbard63,Gutzwiller63}, 
a basic model in 
condensed matter physics due to its relevance for strong electron correlation
phenomena like itinerant magnetism or, for meanwhile ten years, high T$_C$ 
superconductivity \cite{FuldeBuch}. 
In 1980 Kubo \cite{Kubo80} published a high temperature expansion, and 
the nowadays developed cumulant expansions \cite{RKubo62,Becker87} 
and in future the
incremental method \cite{Kladko97} seems to have the potential for a break 
through in the direction towards arbitrary temperatures. 
Nevertheless, at the moment the latter theories 
are mainly elaborated for the ground state properties, 
whereas the finite temperature business was left aside.
Another way, which was shown to be equivalent to the
cumulant technique \cite{Schork92} at least for the ground state, 
is the coupled cluster expansion \cite{CCmethod}. An extension to finite
temperature was given in \cite{Metzner91}. 
In \cite{Espinosa96} an algebraic
approach to operator perturbation theory was presented for zero temperature, 
nevertheless an extension to finite temperatures seems to be possible also 
in this line.
The series expansion for the 
thermodynamical potential can be generated from series expansion of the one 
particle green function, what of course also introduces
(imaginary) time variables. In the following we will develope a series
expansion avoiding this difficulty. 
\section{The perturbation series for the partition sum}
We assume, that the system under consideration is described by a
Hamiltonian $\hamilton$, which can be splitted into two parts, i.e.
\beq
\hamilton=\hamilton_0+\hamilton_1 \, ,
\eeq 
The partitition sum is
\beq
Z&=&Sp \left \{\expo{-\beta(\hamilton_0+\hamilton_1)}  \right \}
\eeq
This can be rewritten in the following form 
\beq
Z&=&Sp \left \{\expo{-\beta\hamilton_0}  \operator{S}(\beta) \right \}
\label{S}
\eeq
with
\beq
\operator{S}(\beta)
&=&\expo{\beta \hamilton_0} \expo{-\beta \hamilton} \label{Smatrix} 
\eeq
The perturbation expansion for the operator $S(\beta)$ given in various
textbooks on quantum statistics is
\beq
\operator{S}(\beta)=\sum_{n=0}^{\infty}(-1)^n
\int_{0}^{\beta}d\beta_1\int_{0}^{\beta_1}d\beta_2 \cdots
\int_{0}^{\beta_{n-1}}d\beta_n
\hamilton_1(\beta_1)\hamilton_1(\beta_2) \cdots \hamilton_1(\beta_n) 
\label{StandardS}
\eeq
with $\hamilton_1(\tau)$ being
\beq
\hamilton_1(\tau)&=&\expo{\tau \hamilton_0}\hamilton_1\expo{-\tau \hamilton_0}
\eeq
It is this (imaginary) time dependence and the multiple time integrals 
which makes the calculation of the perturbation series involved. 
In the following we show that one can get rid of this problem.
For that, we consider both parts of the Hamiltonian as elements of the same 
operator space, and,
therefore, $\hamilton_1$ can be expressed in terms of eigenoperators
of the Liouvillian belonging to $\hamilton_0$. The Liouvillian of
a Hamiltonian is defined as
\beq
\liouville \opA& = &\left [ \hamilton, \opA \right]
\eeq
and an eigenoperator $\opA$  fullfills the eigenvalue equation
\beq
\liouville \opA &=& \lambda_A \opA \, .
\eeq 
The product of two eigenoperators $\opA$ and $\opB$ is also an eigenoperator, with
\beq
\liouville (\opA\opB)&=&(\liouville\opA)\opB+\opA(\liouville\opB)
=\lambda_A\opA\opB+\opA\lambda_B\opB=(\lambda_A+\lambda_B)(\opA\opB)
\eeq
Next, we expand the $\operator{S}$ from eq (\ref{Smatrix}) into a Taylor series
\beq
\operator{S}(\beta)
&=&\sum_{n=0}^{\infty} \frac{\beta^n}{n!} \left[ 
\frac{\partial^n}{\partial \beta^n} \operator{S}(\beta) \right]_{\beta=0}
\eeq
For the first derivative of $\operator{S}$ we find 
\beq
\frac{\partial}{\partial \beta} \operator{S}(\beta)
&=& \operator{a}_{11}(\beta) \operator{S}(\beta)
\eeq
with
\beq
\operator{a}_{11}(\beta)
&=&
\expo{\beta \hamilton_0} \operator{a}_{11} \expo{-\beta\hamilton_0}
= -\expo{\beta \hamilton_0} \hamilton_1 \expo{-\beta\hamilton_0}
\label{a11}
\eeq
The operator $\operator{a}_{11}$ is first order in the perturbation.  
The second derivative is
\beq
\frac{\partial^2}{\partial \beta^2} \operator{S}(\beta)
&=& \left( \frac{\partial}{\partial \beta}\operator{a}_{11}(\beta) 
+ \operator{a}_{11}(\beta)\operator{a}_{11}(\beta) \right ) \operator{S}(\beta)
\nonumber \\
&=&
\left( \operator{a}_{21}(\beta)  + \operator{a}_{22}(\beta)\right)
\operator{S}(\beta)
\eeq
Since the derivation of an $\operator{a}$-operator with respect to $\beta$
does not increase the order of the perturbation $\operator{a}_{21}(\beta)$
remains first order whereas 
$\operator{a}_{22}(\beta)=\operator{a}_{11}^2(\beta)$ is
second order.    
The next derivative is
\beq
\frac{\partial^3}{\partial \beta^3} \operator{S}(\beta)
&=& \left( \frac{\partial}{\partial \beta}\operator{a}_{21}(\beta) 
	 + \frac{\partial}{\partial \beta}\operator{a}_{22}(\beta)
         + (\operator{a}_{21}(\beta)+\operator{a}_{22}(\beta)) \operator{a}_{11}(\beta)
    \right ) \operator{S}(\beta)
\eeq
Collecting together terms of the same order in the perturbation 
yields
\beq
\frac{\partial^3}{\partial \beta^3} \operator{S}(\beta)
&=&
\left( \operator{a}_{31}(\beta)  + \operator{a}_{32}(\beta) 
      + \operator{a}_{33}(\beta) \right) \operator{S}(\beta)
\eeq
Proceeding in this way we can sort the summands contributing to the
expansion of $\operator{S}$ in a table. Since the derivatives have to
be taken at infinite temperature (i.e. at $\beta=0$) we have  
$\operator{S}_{0}=1$. Furthermore, the derivative of an $\operator{a}_{nm}$
with respect to $\beta$ taken at $\beta=0$ is nothing but applying
the Liouvillian to that operator. In table 1 the column index 
gives the order in the perturbation whereas the row index gives the order
in $\beta$.
\begin{table}[hbt]
\begin{center}
\setlength{\unitlength}{1cm}
$
\begin{array}{c|@{\qquad}c@{\hskip 1.5cm}c@{\hskip 1.5cm}c@{\hskip 1.5cm}c}
&1.&2.&3.&4.\\[0.2cm]\hline
&&&&\\
\quad\frac{\beta}{1!}\quad & 
\rnode{a11}{\operator{a}_{11}}&&&\\[1cm]
\frac{\beta^2}{2!} &
\rnode{a21}{\operator{a}_{21}}&\rnode{a22}{\operator{a}_{22}}&&\\[1cm]
\frac{\beta^3}{3!} &
\rnode{a31}{\operator{a}_{31}}&\rnode{a32}{\operator{a}_{32}}& 
\rnode{a33}{\operator{a}_{33}}&\\[1cm]
\frac{\beta^4}{4!} &
\rnode{a41}{\operator{a}_{41}}&\rnode{a42}{\operator{a}_{42}}& 
\rnode{a43}{\operator{a}_{43}}&\rnode{a44}{\operator{a}_{44}}
\end{array}
$
\psset{nodesep=3pt}
\ncline{->}{a11}{a21}\mput*{\tiny $\operator{L}_0$}
\ncline{->}{a11}{a22}\mput*{\tiny $\times\operator{a}_{11}$}
\ncline{->}{a21}{a31}\mput*{\tiny $\operator{L}_0$}
\ncline{->}{a21}{a32}\mput*{\tiny $\times\operator{a}_{11}$}
\ncline{->}{a22}{a32}\mput*{\tiny $\operator{L}_0$}
\ncline{->}{a22}{a33}\mput*{\tiny $\times\operator{a}_{11}$}
\ncline{->}{a31}{a41}\mput*{\tiny $\operator{L}_0$}
\ncline{->}{a31}{a42}\mput*{\tiny $\times\operator{a}_{11}$}
\ncline{->}{a32}{a42}\mput*{\tiny $\operator{L}_0$}
\ncline{->}{a32}{a43}\mput*{\tiny $\times\operator{a}_{11}$}
\ncline{->}{a33}{a43}\mput*{\tiny $\operator{L}_0$}
\ncline{->}{a33}{a44}\mput*{\tiny $\times\operator{a}_{11}$}
\par
\parbox{10cm}{\caption{Operators $\operator{a}_{nm}$ 
of the same order in the perturbation
are depicted in the columns. The rows give the operators contributing to the
same power of the temperature. The arrows symbolize the recursive connection
between the operators.  A vertical arrow stands for application of
$\operator{L}_0$ and a slanted arrow for post-multiplying 
with $\operator{a}_{11}$. }}
\end{center}
\end{table}
Thus by resorting the sum one finds
\beq
\operator{S}(\beta)&=&
1 + \operator{S}_1(\beta)+\operator{S}_2(\beta)+\operator{S}_3(\beta)+ \cdots
\eeq
with
\beq
\operator{S}_m(\beta)&=&
\sum_{n=m}^{\infty} \frac{\beta^n}{n!} \operator{a}_{nm}
\label{Sasum}
\eeq
It is easy to show that for the operators $\operator{a}_{mn}$ the
following recursion relations hold
\beq
\operator{a}_{mm}&=&\operator{a}_{11}^m   \\
\operator{a}_{n1}&=&\liouville_0^{n-1} \operator{a}_{11} \\
\operator{a}_{nm}&=&\liouville_0 \operator{a}_{n-1\,m} 
+ \operator{a}_{n-1\,m-1} \operator{a}_{11}            \label{recursion}
\eeq
By help of the above equations (\ref{recursion}) the following formula 
results
\beq  
\operator{a}_{nm} &=&\liouville_0^{n-m}\operator{a}_{mm} +
\sum_{k=0}^{n-1-m} \liouville_0^{k}(\operator{a}_{n-1-k\,m-1} 
\operator{a}_{11}) \label{reduction} 
\eeq
The recursion relations allow to find closed formula for the
m-th order contribution to the partition sum. We will show this
explicit for the second order contribution and the third order 
contribution respectively.
Summing the second and the third column in table 1 respectively yields for the operators
$\operator{S}_{2}$ and $\operator{S}_{3}$  
\beq
\operator{S}_2&=&
\sum_{n=2}^{\infty} \frac{\beta^n}{n!} \operator{a}_{n2}
\qquad ; \qquad
\operator{S}_3=
\sum_{n=3}^{\infty} \frac{\beta^n}{n!} \operator{a}_{n3}
\eeq
Iterating eq (\ref{reduction}) one gets for the operators $\operator{a}_{n2}$
\beq
\operator{a}_{n2}&=&\delta_{n,2}\operator{a}_{22}
+\delta_{n>2} \left \{\liouville_0^{n-2}\operator{a}_{22} +
\sum_{k=0}^{n-3} \liouville_0^{k}(\operator{a}_{n-1-k\,1} \operator{a}_{11})
\right \}    \nonumber          \\
&=&\delta_{n,2}\operator{a}_{22}
+
\delta_{n>2} \left \{\liouville_0^{n-2} \operator{a}_{22} +
\sum_{k=0}^{n-3} \liouville_0^{k}\left [ (\liouville_0^{n-1-k} \operator{a}_{11}
)
\operator{a}_{11}) \right ]
\right \}
\label{opan2}
\eeq
Here the shortcut $\delta_{n>m}$ says that n has to be greater than m.
For the third order results
\beq
\operator{a}_{n3}&=&\delta_{n,3} \, \operator{a}_{33}
+
\delta_{n>3} \,
\liouville_0^{n-3} \operator{a}_{33} + \nonumber \\  
&&\hspace{1cm}+\delta_{n>3}
\left \{
\sum_{k=0}^{n-4}\liouville_0^{k}
\left(
\liouville_0^{n-3-k}\operator{a}_{22}+ 
\sum_{l=0}^{n-4-k} \liouville_0^l
\left[ 
(\liouville_0^{n-3-k-l} \operator{a}_{11}) \operator{a}_{11} 
\right] 
\operator{a}_{11} \right) \right \} \label{opan3}
\eeq
One finds with eqs (\ref{S},\ref{Sasum}) for the m-th order
contribution to the partition sum 
\beq
Z_m/Z_0&=&Sp \left \{\expo{-\beta\hamilton_0}  
\sum_{n=m}^{\infty} \frac{\beta^n}{n!} \operator{a}_{nm}
 \right \}=\sum_{n=m}^{\infty} \frac{\beta^n}{n!}\expect{\operator{a}_{nm}}
\label{Sexpect}
\eeq 
what can be rewritten by help of eq (\ref{reduction}) 
\beq
Z_m/Z_0&=&\sum_{n=m}^{\infty} \frac{\beta^n}{n!}
\left \{ \expect{\opL_0^{n-m}\operator{a}_{mm}}
+\delta_{n>m}\sum_{k=0}^{n-1-m}\expect{\opL_0^{k}(\operator{a}_{n-1-k, m-1}
\operator{a}_{11})} \right \}
\eeq
Since for every operator $\opA$  holds
\beq
\expect{\opL_0^{n-m}\opA}&=&\delta_{nm}\expect{\opA}
\eeq
we find
\beq
Z_m/Z_0&=& \frac{\beta^m}{m!}  \expect{\operator{a}_{mm}}
+ \sum_{n=m+1}^{\infty} \frac{\beta^n}{n!}\expect{\operator{a}_{n-1, m-1}
\operator{a}_{11}} 
\label{Firstiter}
\eeq
the next recursion step yields
\beq
Z_m/Z_0&=& \frac{\beta^m}{m!}  \expect{\operator{a}_{mm}}
+ \sum_{n=m+1}^{\infty} \frac{\beta^n}{n!}
\expect{(\opL_0^{n-m}\operator{a}_{m-1, m-1})
\operator{a}_{11}}  \nonumber \\
&&
+ \sum_{n=m+1}^{\infty} \frac{\beta^n}{n!}
\sum_{k=0}^{n-1-m}\expect{\left(\opL_0^k(\operator{a}_{n-2-k,
m-2}\operator{a}_{11})\right)\operator{a}_{11}}
\label{Seconditer}
\eeq
The recursion ends when the operators are reduced to products 
of $\operator{a}_{11}$.
Thus we get for the second order
from eq (\ref{Firstiter}) 
\beq
Z_2/Z_0&=&
 \frac{\beta^2}{2!}  \expect{\operator{a}_{22}}
+ \sum_{n=3}^{\infty} \frac{\beta^n}{n!}\expect{\operator{a}_{n-1 1}
\operator{a}_{11}} 
\nonumber \\
&=&
\frac{\beta^2}{2!}  \expect{\operator{a}_{11}^2}
+ \sum_{n=3}^{\infty} \frac{\beta^n}{n!}\expect{(\opL_0^{n-2}\operator{a}_{11})
\operator{a}_{11}} 
\label{Z2a11}
\eeq     
and for the third order
\beq
Z_3/Z_0&=&
 \frac{\beta^3}{3!}  \expect{\operator{a}_{33}}
+ \sum_{n=4}^{\infty} \frac{\beta^n}{n!}
\expect{\left(\opL_0^{n-3}\operator{a}_{22}\right)
\operator{a}_{11}} 
+ \sum_{n=4}^{\infty} \frac{\beta^n}{n!}\sum_{l=0}^{n-4}
\expect{\left(\opL_0 \left( \opL_0^{n-3-l}\operator{a}_{11}\right ) 
\operator{a}_{11}\right)\operator{a}_{11}} \nonumber \\
&=&
 \frac{\beta^3}{3!}  \expect{\operator{a}_{11}^3}
+ \sum_{n=4}^{\infty} \frac{\beta^n}{n!}
\expect{\left(\opL_0^{n-3}\operator{a}_{11}^2\right)
\operator{a}_{11}} 
+ \sum_{n=4}^{\infty} \frac{\beta^n}{n!}\sum_{l=0}^{n-4}
\expect{\left(\opL_0 \left( \opL_0^{n-3-l}\operator{a}_{11}\right ) 
\operator{a}_{11}\right)\operator{a}_{11}}
\label{Z3a11}
\eeq
Next, the perturbation $\hamilton_1$ is decomposed into eigenoperators of
$\liouville_0$. For that we use the natural basis of the Liouville space
formed by the dyades constructed from the eigenstates of $\hamilton_0$.
In this representation $\hamilton_0$ is diagonal and $\hamilton_1$ is
chosen purely offdiagonal, what is always possible. In case that 
$\hamilton_1$ contains components which are diagonal they are added to 
$\hamilton_0$.      
\beq
\hamilton_0=\summe{\nu} \varepsilon_{\nu} \mket{\nu}\mbra{\nu} 
\qquad \mbox{with} \qquad
\hamilton_0 \mket{\nu}=\varepsilon_{\nu} \mket{\nu} \,.
\eeq
and
\beq
\hamilton_1=\summe{\mu} \summe{\nu} V_{\mu\nu} \mket{\mu}\mbra{\nu} 
\label{VinLform}
\eeq
The basis operators $\mket{\mu}\mbra{\nu}$ are eigenoperators of the
Liouvillian $\liouville_0$ following   
\beq
\liouville_0 \mket{\mu}\mbra{\nu}=\lambda_{\mu\nu}\mket{\mu}\mbra{\nu} 
\qquad \mbox{with} \qquad
\lambda_{\mu\nu}=\varepsilon_{\mu}-\varepsilon_{\nu}   \label{eveqn}
\eeq
For the operator $\operator{a}_{11}$ we get
\beq  
\operator{a}_{11}=\summe{\mu} \summe{\nu} A_{\mu\nu} \mket{\mu}\mbra{\nu} 
\qquad \mbox{with} \qquad
A_{\mu\nu} = -V_{\mu\nu}
\eeq
By insertion of the above expression into eqs (\ref{Z2a11},\ref{Z3a11}) and 
using the
linearity of the Liouvillian the result for
the second order contribution to the partition sum is
\beq
Z_2&=& \summe{\mu}\summe{\nu} \expo{-\beta\varepsilon_{\mu}}  V_{\mu\nu}V_{\nu\mu}
	f_2(\lambda_1) \label{Z2}
\eeq
with
\beq
f_2(\lambda_1)&=& \frac{\beta^2}{2!}+\frac{E_2(\lambda_1)}{\lambda_1^2}
\eeq
and the third order contribution reads as
\beq
Z_3&=& -\summe{\mu}\summe{\nu} \summe{\alpha}
\expo{-\beta\varepsilon_{\mu}}  V_{\mu\nu}V_{\nu\alpha}V_{\alpha\mu}
	f_3(\lambda_1,	\lambda_2)
\eeq
with
\beq
f_3(\lambda_1,\lambda_2)&=& \frac{\beta^3}{3!}+
\frac{1}{\lambda_2-\lambda_1}
\frac{E_3(\lambda_1)}{\lambda_1^2}
+
\frac{1}{\lambda_1-\lambda_2}
\frac{E_3(\lambda_2)}{\lambda_2^2} 
\label{f3}
\eeq
Here we introduced the functions $E_m(\lambda)$, which are determined
according to
\beq
E_m(\lambda)&=&
\expo{\beta \lambda}-\sum\limits_{n=0}^{m}\frac{ (\beta \lambda)^n}{n!} 
\label{Em}
\eeq
Furthermore we abbreviated
\beq
\lambda_1&=&\lambda_{\mu\mu'}\\
\lambda_2&=&\lambda_{\mu\mu'}+\lambda_{\nu\nu'}
\eeq
In the same way we proceeded to higher orders in the perturbation.
We calculated the coefficients to the eighth order terms,
what one finds for the m$^{th}$-order contribution to the partition
sum is
\beq
Z_m=(-1)^m\summe{\mu_1} \cdots \summe{\mu_m}\expo{-\beta\varepsilon_{\mu_1}}
V_{\mu_1\mu_2}V_{\mu_2\mu_3} \cdots V_{\mu_{m-1}\mu_m}V_{\mu_m\mu_1} 
f_m(\lambda_1,\dots,\lambda_{m-1})   \label{series}
\eeq
with
\beq
f_m(\lambda_1,\dots,\lambda_{m-1})&=&
\frac{\beta^m}{m!}+\sum_{i=1}^{m-1} \frac{\lambda_i^{m-2}}
{\prod_{j \neq i}(\lambda_j-\lambda_i)} \frac{E_m(\lambda_i)}{\lambda_i^m}
\label{fm}
\eeq
and
\beq
\lambda_k&=&\lambda_{\mu_1\mu_2}+\lambda_{\mu_2\mu_3} + \cdots 
+\lambda_{\mu_k\mu_{k+1}} =\varepsilon_{\mu_1}-\varepsilon_{\mu_k}
\eeq
We want to emphasize, that the functions 
$f_m(\lambda_1, \cdots ,\lambda_{m-1})$ 
remain finite for any number of vanishing $\lambda_i$, and, therefore, 
the case that the unperturbed Hamiltonian has a degenerated spectrum 
is included.
Furthermore, it remains finite for all values of $\beta$ . This becomes
immediately clear, if one looks at the structure of the operators
$\operator{a}_{nm}$,
which do not contain any denominators,  
e.g. look at $\operator{a}_{n2}$ and $\operator{a}_{n3}$
as given in eqs (\ref{opan2}) and (\ref{opan3}) respectively.       
Eq (\ref{series}) provides a compact form for the perturbation series of the
partition sum which can be tested easily for simple systems. As will be 
shown in the following the functions
$f_m(\lambda_1,\cdots,\lambda_{m-1})$ are nothing but the result of the
m-1 integrations necessary in standard perturbation theory. 
Thus we did this integrals to infinite order. 
Starting from eq (\ref{StandardS}) we can insert the perturbation in the
form given in eq (\ref{VinLform}), what yields
\beq
Z(\beta)&=&\sum_{m=0}^{\infty}(-1)^m
\summe{\nu_1}\summe{\nu_2}\cdots\summe{\nu_{m}}
\expo{-\beta\varepsilon_{\nu_1}}
V_{\nu_1,\nu_2}V_{\nu_2,\nu_3}\cdots V_{\nu_m,\nu_1} \times \\
&\times&\int\limits_{0}^{\beta}d\beta_1\int\limits_{0}^{\beta_1}d\beta_2
 \cdots 
\int\limits_{0}^{\beta_{m-1}}\!d\beta_m
\expo{\beta_1(\varepsilon_{\nu_1}-\varepsilon_{\nu_2})}
\expo{\beta_2(\varepsilon_{\nu_2}-\varepsilon_{\nu_3})}
\cdots
\expo{\beta_m(\varepsilon_{\nu_m}-\varepsilon_{\nu_1})}\nonumber  \label{tauint}
\eeq
due to the eigenvalue equation
of the unperturbed Hamiltonian for the partition sum.
It is obvious that the result of the m integrations have to be the
same as the functions introduced in eq (\ref{fm}).
Taking into account the definitions of the $\lambda_i$ one finds
\beq
\varepsilon_{\nu_1}-\varepsilon_{\nu_2}&=&
\lambda_1\\
\varepsilon_{\nu_i}-\varepsilon_{\nu_{i+1}}&=&
-\lambda_{i-1}+\lambda_{i} \qquad \mbox{with}\qquad 1<i<m
\eeq
and, therefore,
\beq
f_m(\lambda_1,\dots,\lambda_{m-1})\!\!&=&\!\!
\int\limits_{0}^{\beta}\!d\beta_1\int\limits_{0}^{\beta_1}\!d\beta_2 \cdots
\!\!\int\limits_{0}^{\beta_{m-1}}\!\!\!d\beta_m
\expo{\beta_1\lambda_1}
\expo{\beta_2(-\lambda_1+\lambda_2)}
\cdots
\expo{\beta_m(-\lambda_{m-1} )} \label{fmviatauint}
\eeq
These integrals were calculated by help of symbolic computer programs.
Up to order five ($2^5$ terms), it was possible to bring the results
symbolically to the forms given in eq. (\ref{fm}) by help of the tools, the
symbolic computer languages provide.  For the
 sixth, seventh, and eighth order term we proved it by numerical calculation,
for arbitrary sets of $\lambda$'s, and found reasonable coincidence 
within the numerical error. A different way to do the integrals is 
by help of Laplace's transformation \cite{John97}. This results in a form
of $f_m$ which seems at the first glance different. Nevertheless it can be
shown to be equivalent to eq (\ref{fm}). The calculation is given in the 
appendix D.
For comparison with the literature, where results are
given for the series expansion of the grand potential, we have to 
express the perturbation series for the grand potential $F$ via the
perturbation series of the partition sum. One finds easily
\beq
\beta F&=&\beta F_0+\beta F_2+\beta F_3+ \beta F_4 ... \nonumber \\
       &=& -\log Z_0-\frac{Z_2}{Z_0} -\frac{Z_3}{Z_0} 
- \left (\frac{Z_4}{Z_0}-\frac{Z_2^2}{Z_0^2} \right )+ ...
\eeq
This expression may be compared to the usual perturbation theory
for the grand potential \cite{Abrikosov}
\beq
\beta F&=&\beta F_0-\sum_{n=2}^{\infty}(-1)^n
\int_{0}^{\beta}d\beta_1\int_{0}^{\beta_1}d\beta_2 \cdots
\int_{0}^{\beta_{n-1}}d\beta_n
\expect{\hamilton_1(\beta_1)\hamilton_1(\beta_2) 
\cdots \hamilton_1(\beta_n)}^c 
\label{StandardF}
\eeq
here $\expect{\dots}^c$ indicates that cumulants \cite{RKubo62} have to be
calculated.
Due to the linearity of cumulants we can separate the $\tau$-integrations
what yields together with eq (\ref{fmviatauint})
\beq
\beta F&=&\beta F_0
-\sum_{n=2}^{\infty}(-1)^n
 \summe{\mu_1} \cdots \summe{\mu_n}\expo{-\beta\varepsilon_{\mu_1}}
f_n(\lambda_1,\dots,\lambda_{n-1}) \times \\
&&\hspace{2cm}\times \expect{V_{\mu_1\mu_2}V_{\mu_2\mu_3} \cdots 
                      V_{\mu_{n-1}\mu_n}V_{\mu_n\mu_1}}^c 
\nonumber  
\label{SeriesFc}
\eeq
Thus it is sufficient to calculate cumulants, i.e. the linked graphs.
\section{Calculating the expectation values}
What remains to calculate is the expectation value of eigenoperator products, 
being of the form
\beq
\expect{\opA \opB}&=&\frac{1}{Z_0}
Sp \left \{ \expo{-\beta \hamilton_0} \opA \opB \right \}
\eeq
Here $\opA$ is an arbitrary eigenoperator of $\liouville_0$ with eigenvalue 
$\lambda_A$ and $\opB$
any other operator, $\langle\cdots\rangle$ is again the expectation value with
respect to the unperturbed Hamiltonian .
Since cyclic permutations under a trace do not alter the expectation value
we find
\beq
Sp\left \{ \expo{-\beta \hamilton_0}\opA\opB \right \}
&=&
Sp\left \{ \left( \expo{-\beta \liouville_0} 
\opA \right ) \expo{-\beta \hamilton_0} \opB \right \} 
=\,\expo{-\beta \lambda_A } Sp \left \{  
\expo{-\beta \hamilton_0} \opB \opA \right \}\nonumber \\
&=&\expo{-\beta \lambda_A }Sp\left \{  
\expo{-\beta \hamilton_0} [ \opB, \opA]_{+} \right \}
-\expo{-\beta \lambda_A }Sp\left \{  
\expo{-\beta \hamilton_0} \opA \opB \right \}
\eeq
Solving for $\expect{\opA\opB}$ yields
\beq
\expect{\opA \opB}&=&\frac{1}{\expo{\beta\lambda_A}+1}
\expect{[\opA,\opB]_{+}}  \label{anticommutatorfall} 
\eeq
If $\opB$ is an eigenoperator instead of $\opA$ one finds in the
same way
\beq
\expect{\opA \opB}&=&\frac{1}{\expo{-\beta\lambda_B}+1}
\expect{[\opA,\opB]_{+}}
\eeq 
Of course one can get these results also via the standard Green function 
technique which becomes extremly simple for eigenoperators. 
This is shown in appendix A.
If both $\opA$ and $\opB$ are eigenoperators, than one has
\beq
Sp\left \{ \expo{-\beta \hamilton_0}\opA\opB \right \}
&=&
Sp\left \{ \expo{-\beta \hamilton_0}\opA
\opB \expo{\beta \hamilton_0}
\expo{-\beta \hamilton_0}  \right \}  \\
&=&
Sp\left \{ 
\left ( \expo{-\beta \liouville_0}\opA \opB \right )
\expo{-\beta \hamilton_0}  \right \}
=\expo{-\beta (\lambda_A+\lambda_B)}
Sp\left \{ \expo{-\beta \hamilton_0} \opA \opB \right \}\nonumber
\eeq
It follows that either the expectation values $\expect{\opA\opB}$ and 
$\expect{\opB\opA}$ vanish or the equation
\beq
\lambda_B=-\lambda_A   \qquad \mbox{ for $\opA$, $\opB$ eigenoperators of 
$\liouville_0$ } \label{agleichb}
\eeq
holds.
In some models, e.g. within the Hubbard model, 
the operators $\opA$ and $\opB$ may be both fermionic and bosonic.
If both operators are fermionic, than the anticommutator is suitable, since
the number of operators will be reduced.
In case that at least one of the operators $\opA$ or $\opB$ is bosonic,
than its more convenient to work with the commutator. The related formula
can be derived easily from eq (\ref{anticommutatorfall}) to be
\beq
\expect{\opA \opB}&=&\frac{1}{1-\expo{ \beta\lambda_A}}
\expect{[\opA,\opB]}  \label{commutatorfall} 
\eeq
Introducing a "parity function" $P(\opA |\opB)$ being an odd integer if both
operators are fermionic and even else one can unite 
eqs (\ref{anticommutatorfall},\ref{commutatorfall}) in the following
way
\beq
\expect{\opA \opB}&=&\frac{1}{1-(-1)^{P(A|B)}\expo{ \beta\lambda_A}}
\expect{\opA\opB - (-1)^{P(A|B)}\opB\opA}  \label{unitedfall} 
\eeq
In the above form the number of operators will be always reduced.  
It is of some interest to discuss the case when
the denominator in eq (\ref{commutatorfall}) vanishes. This may happen
if the temperature goes to infinity. Since there is no reason that
the expectation value $\expect{\opA\opB}$ becomes infinite, we have to
conclude, that the commutator becomes zero, what represents the classical 
limit. More interesting is the case that the eigenvalue $\lambda_A$ is
zero. But this implies the following statement:
If the operator $\opA$ is an eigenoperator to the eigenvalue zero, then
it is diagonal or at least nondiagonal only between degenerated states. 
Indeed from expanding the operator
$\opA$ in the basis of $\liouville_0$ and applying the eigenvalue equation
follows
\beq
\liouville_0 \opA 
&=& 
\liouville_0 \summe{\mu}\summe{\nu} A_{\mu\nu}\mket{\mu} \mbra{\nu}
= \summe{\mu}\summe{\nu} A_{\mu\nu} 
(\varepsilon_{\mu}-\varepsilon_{\nu})\mket{\mu} \mbra{\nu} = \operator{0}
\eeq
Taking the matrix elements yields
\beq
0 &=&A_{\mu\nu}(\varepsilon_{\mu}-\varepsilon_{\nu})
\eeq
what proves the statement.    
Of course the same holds if $\opB$ belongs to the eigenvalue zero.
Furthermore, from eq (\ref{agleichb}) one can derive the following statement:
If both operators $\opA$ and $\opB$ respectively are eigenoperators and
at least one of them has the eigenvalue zero, then the expectation value
vanishes or both eigenvalues are zero.
Thus, for all nonvanishing expectation values both operators have to
be diagonal, if one is. 
In the latter case the two operators commute. One can also prove the following
statements, in some sense reverse to the above said.
If two eigenoperators $\opA$ and $\opB$ 
anticommute their expectation value $\expect{\opA\opB}$ vanishes. 
This follows from equation
(\ref{commutatorfall}) 
\beq
\expect{\opA\opB}&=&  
\frac{1}{1-\expo{\beta\lambda_A}}\expect{[\opA,\opB]}
\,=\,
\frac{2}{1-\expo{\beta\lambda_A}}\expect{\opA\opB}
\eeq
but this means $\expect{\opA\opB}=0$. 
Furthermore, if two eigenoperators $\opA$ and $\opB$ 
commute their expectation value $\expect{\opA\opB}$ 
either vanishes or both eigenvalues have to be zero.
Indeed, we get from eq (\ref{anticommutatorfall})
\beq
\expect{\opA\opB}&=&  
\frac{1}{1+\expo{\beta\lambda_A}}\expect{[\opA,\opB]_{+}}
\,=\,
\frac{2}{1+\expo{\beta\lambda_A}}\expect{\opA\opB}
\eeq 
Again one possible solution is $\expect{\opA\opB}=0$, but
if we demand that $\expect{\opA\opB} \neq 0$ then necessarily
$\lambda_A =0 $ holds. Besides $\lambda_A$ also $\lambda_B$ has
to vanish which follows from eq (\ref{agleichb}).
Thus the number of operators inside an expectation value can
be reduced by help of eq (\ref{unitedfall}) till all remaining operators are
commuting and diagonal. 
For illustration, let us assume for the moment both $\opA$ and $\opB$ 
to be simple basis operators of the form
\beq
\opA&=&\mket{\mu}\mbra{\mu'} \qquad \mbox{with} \qquad 
\lambda_A=\varepsilon_{\mu}-\varepsilon_{\mu'} \\
\opB&=&\mket{\nu}\mbra{\nu'} \qquad \mbox{with} \qquad 
\lambda_B=\varepsilon_{\nu}-\varepsilon_{\nu'}
\eeq
then we find 
\beq
\langle [\opA,\opB]_{+}\rangle&=&
\left (\expo{-\beta\varepsilon_{\mu}}+\expo{-\beta\varepsilon_{\nu}}\right )
\frac{\delta_{\mu\nu'}\delta_{\nu\mu'}}{Z_0} \\
\langle \opA\opB\rangle&=&
\frac{ \expo{-\beta\varepsilon_{\mu}}+\expo{-\beta\varepsilon_{\nu}}}
{\expo{\beta(\varepsilon_{\mu}-\varepsilon_{\nu})}+1}
\frac{\delta_{\mu\nu'}\delta_{\nu\mu'}}{Z_0}
=\frac{\expo{-\beta\varepsilon_{\mu}}}{Z_0}\delta_{\mu\nu'}\delta_{\nu\mu'}
\eeq
For arbitrary non-diagonal operators 
\beq
\opA=\summe{\mu\neq\nu} A_{\mu\nu}\mket{\mu}\mbra{\nu}
&\qquad \mbox{and} \qquad &
\opB=\summe{\mu'\neq\nu'} B_{\mu'\nu'}\mket{\mu'}\mbra{\nu'}
\eeq
one finds 
\beq
\expect{\opA\opB}&=&
\summe{\mu\neq\nu}\frac{\expo{-\beta\varepsilon_{\mu}}}{Z_0}
A_{\mu\nu}B_{\nu\mu}
\eeq
If we take both $\opA$ and $\opB$ to be the perturbation we get together
with eq (\ref{Zm}) the result for $Z_2$ in accordance with eq (\ref{Z2}).
In appendix C we show how eq (\ref{unitedfall}) together with the
statements given above can be utilized to evaluate
expectation values within the Hubbard model in a systematic manner.
\section{The one band Hubbard model}
We will demonstrate it for the Hubbard model,
being the most simple lattice fermion model taking into account electron
electron interaction. In the context of strong electron correlation especially
the perturbation expansion
around the atomic limit is of interest. Therefore, we focus the discussion 
to that case.
The Hamiltonian of the Hubbard model for a grand canonical ensemble is 
\beq
\hamilton &=& \hamilton_0 + \hamilton_1
\eeq
with
\beq
\hamilton_0&=& \summe{i \sigma} \left(
\frac{U}{2}\nsigma{i}\noperator{i-\sigma}-(\mu+\sigma h_i) \nsigma{i} \right )\\
\label{hamilton}
\hamilton_1&=& t \sumijs \cplus{i \sigma}\cminus{j \sigma}
\eeq
Here $\cplus{i \sigma}$ and $\cminus{i \sigma}$ are the creation and 
destruction operators
in Wannier representation. 
The chemical potential $\mu$  and the magnetic 
field in z-direction are introduced
to take the effects of doping and applying external magnetic fields into 
account.
The model has two exact solveable limits, i.e. the band limit with
$U=0$ and the  atomic limit, where $t=0$ holds.
Since we are interested in the large U limit, we use as unperturbed Hamiltonian
$\hamilton_0$, the so called   
atomic limit of the Hubbard model.
Within this limit the electrons are at N independent lattice sites and
the partition sum factorizes. The eigenoperators of the related Liouvillian
$\opL_0$ are the so called Hubbard operators \cite{Hubbard65} and 
products of them.
Since the multisite Hilbertspace and also the related Liouville space are 
the direct product of the N single site spaces, it is enough
to restrict the discussion to one lattice site indexed by i for the moment. 
For one lattice site the atomic limit Hamiltonian is
\beq
\hamilton_i=U \nup{i}\ndown{i}- h_i(\nup{i}-\ndown{i})-\mu(\nup{i}+\ndown{i})
\label{local_h}
\eeq 
and its Hilbertspace is spanned by the four eigenstates:
\beq
\ket{i,0}&& \qquad \mbox{if the lattice site is empty, 
          }    \\
\ket{i,u}&& \qquad \mbox{if the lattice site is occupied with a spin-up electron, 
          } \\
\ket{i,d}&& \qquad \mbox{if the lattice site is occupied with a spin-down electron, 
          } \\
\ket{i,2}&& \qquad \mbox{if the lattice site is occupied with two electrons. 
          }. 
\eeq
The related eigenvalue equations are
\beq 
\hamilton_i \ket{i,\mu}=\varepsilon_{i\mu} \ket{i,\mu}  
\qquad \mbox{with} \qquad \mu,\nu \in \{0,u,d,2\} 
\eeq
and
\beq
\varepsilon_{i0} = 0 \quad , \quad
\varepsilon_{i\sigma} = -\sigma h_i-\mu \quad ,\quad
\varepsilon_{i2} = -2\mu+U \label{epsilonisigma}
\eeq
From these states one can construct the natural basis 
of the related operator space.
This was also first done by Hubbard \cite{Hubbard65} indicating the
basis operators by $X^{\mu\nu}_i=\ket{i,\mu}\bra{i,\nu}$. The eigenvalue 
equation for the basis operators is
\beq
\opL_i X^{\mu\nu}_i&=&\lambda^{\mu\nu}_i X^{\mu\nu}_i  \qquad \mbox{with} 
\qquad \lambda^{\mu\nu}_i = \varepsilon_{i\mu} -\varepsilon_{i\nu}  \quad .
\eeq                             
For a detailed discussion of the physics of these operators and a related 
diagrammatic technique I refer to the book of Isjumov and Skrjabin 
\cite{Isjumov87} and the references therein. 
The fermion creation and destruction operators may be expressed 
via the basis operators according to   
\beq                     
\csigma{i}&=&X^{0\sigma}_i+\sigma X^{-\sigma 2}_i 
\label{cviax}  \\
\cpsigma{i}&=&X^{\sigma 0}_i+\sigma X^{2-\sigma}_i 
\label{cpviax}\\
\nsigma{i}&=&X^{\sigma\sigma}_i+X^{22}_i
\eeq
and for the atomic limit Hamiltonian playing the role of $\hamilton_0$ in the
perturbation theory, we get 
\beq
\hamilton_0&=&\summe{i\sigma} \left ( (\frac{U}{2}-\mu) X^{22}_i-
(\mu+\sigma h_i) X^{\sigma\sigma}_i \right )
\eeq
The related partition sum is  
\beq
Z_0
&=&\prod_i z_i \qquad \mbox{with} \qquad z_i=
1+\expo{\beta(\mu+h_i)}+\expo{\beta(\mu-h_i)}+\expo{\beta(2\mu-U)} 
\label{Z0}
\eeq
The partition sum factorizes into the product of N single site
partition sums. By help of eqs (\ref{cviax},\ref{cpviax}) the hopping part 
may also be rewritten
in terms of single site basis operators
\beq
\hamilton_1&=& t  \sumijs \left ( 
X_i^{\sigma 0} X_j^{0 \sigma} +
\sigma X_i^{\sigma 0} X_j^{-\sigma 2}+
\sigma X_i^{2 -\sigma} X_j^{0 \sigma}+
X_i^{2 -\sigma} X_j^{-\sigma 2} \right )
\eeq
Using the eigenvalue equation for the X-operators yields
\beq
\liouville_0 X_i^{\mu\nu}X_j^{\mu'\nu'}&=&
(\lambda_i^{\mu\nu}+\lambda_j^{\mu',\nu'}) X_i^{\mu\nu}X_j^{\mu'\nu'}
\eeq
Therefore, the operator $\operator{a}_{11}$  defined in eq (\ref{a11}) 
takes the following form   
\beq
\operator{a}_{11}&=-t&\summe{r}\sumijs \opA^r_{ij\sigma}
\label{Hubbarda11}
\eeq
Here I used the abbreviations
\beq
{\opA}^1_{ij\sigma}&=& X_i^{\sigma 0} X_j^{0 \sigma} \label{A}\\
{\opA}^2_{ij\sigma}&=& \sigma X_i^{\sigma 0} X_j^{-\sigma 2}\label{B} \\
{\opA}^3_{ij\sigma}&=& \sigma X_i^{2 -\sigma} X_j^{0 \sigma}\label{C} \\
{\opA}^4_{ij\sigma}&=& X_i^{2 -\sigma} X_j^{-\sigma 2}\label{D} 
\eeq
What remains, is to calculate the matrix elements of $\operator{a}_{11}$.
The straight forward way is to define an eigenstate of $\hamilton_0$ by help 
of the X-operators in the following way
\beq
\mket{\{\mu\}}&=&\prod\limits_{l=1}^{N} X_l^{\mu_l0}\ketvacuum
\eeq
and to calculate the matrix elements. This lengthy calculation is interesting
from a pedagogical point of view and we shift it to the appendix C. 
Here we adopt another way, starting from eq (\ref{Z2a11}). 
Insertion of operator $\operator{a}_{11}$ in the form given in 
eq (\ref{Hubbarda11}) yields for the
Hubbard model 
\beq
Z_2/Z_0&=&
t^2\summe{r}\sumijs\summe{r'}\summe{k\neq l \sigma'}
\left ( \frac{\beta^2}{2!}  + \sum_{n=3}^{\infty} \frac{\beta^n}{n!}
(\lambda_{ij\sigma}^r)^{n-2}
 \right )\expect{ \opA_{ij\sigma}^r \opA_{kl \sigma'}^{r'} } \nonumber \\
&=&
t^2\summe{r}\sumijs\summe{r'}\summe{k\neq l
\sigma'}f_2(\lambda_{ij\sigma}^r)
\expect{\opA_{ij\sigma}^r\opA_{kl\sigma'}^{r'}}
\label{Z2Hubbard}
\eeq
The third order term becomes with eq (\ref{Z3a11})
\beq
Z_3/Z_0&=&-t^3
\summe{r}\sumijs\summe{r'}\summe{i'\neq j' \sigma'}
\summe{r''}\summe{i''\neq j'' \sigma''}
f_3(\lambda_{ij\sigma}^r,\lambda_{ij\sigma}^r+\lambda_{i'j'\sigma'}^{r'})
\expect{\opA_{ij\sigma}^r\opA_{i'j'\sigma'}^{r'}\opA_{i''j''\sigma''}^{r''}}
\eeq
Proceeding in the same way as before yields for 
contribution of order $t^m$
\beq
Z_m/Z_0&=&(-t)^m\summe{x_1} \cdots \summe{x_m}f_m(\lambda_1, \dots , 
\lambda_{m-1})
\expect{\opA_{x_1}\cdots \opA_{x_m}} \label{Zm}
\eeq
Here $x_i$ abbreviates the set of indices r, i, j, and $\sigma$ of the
operator $\opA_{ij\sigma}^r$. The meaning of $\lambda_1$, $\lambda_2$, 
and so on has a little bit changed with respect to eq (\ref{series}), i.e.
\beq
\lambda_1&=&\lambda_{x_1}=\lambda_{ij\sigma}^r \nonumber \\
\lambda_2&=&\lambda_{x_1}+\lambda_{x_2}\nonumber \\
\vdots\nonumber \\
\lambda_k&=&\sum_{i=1}^k \lambda_{x_i}
\eeq
The remaining task is to calculate the expectation values in eq (\ref{Zm}). 
Since all the X-operators appearing in the perturbation are nondiagonal, all
expectation values containing unpaired operators vanish. Thus we have to
take into account all possible systems of paired X-operators, where a factor -1
has to be included, if the number of commutations necessary to make all pairing
X operators neighbours is odd, a factor +1 otherwise. This is nothing but
Wick's theorem. Of course most of these remaining expectation values vanish
also, since not every X-operator fits to each other and the pairing of two 
operators may be on-site non-diagonal. Nevertheless after the first pairing
step every expectation value containing 2m, what is two times the order in t, 
X-operators disintegrates into a finite series 
of expectation values containing m X-operators. 
In case that the lattice site indices of all remaining X-operators
are different one from each other, we have to take into account all terms
containing diagonal X-operators. In case that some of the lattice site indices
are equal we have to contract them again. This way one can systematically find
all nonvanishing contributions to an m-order expectation value. 
The second order expectation value is
\beq
\expect{\opA_{ij\sigma}^r\opA_{mn\sigma'}^{r'}}
&=&
\expect{X_i^{\alpha_i\alpha'_i}X_j^{\xi_j\xi'_j} 
        X_m^{\gamma_m\gamma'_m}X_n^{\eta_n\eta'_n}}
\label{XXXX}
\eeq
Due to the summation restriction $i\neq j$ and $m\neq n$ two
contributions remain
\beq
\expect{X_i^{\alpha_i\alpha'_i}X_j^{\xi_j\xi'_j} 
        X_m^{\gamma_m\gamma'_m}X_n^{\eta_n\eta'_n}}
&=& \\
&&\hspace{-3cm}=-\expect{X_i^{\alpha_i\gamma'_m}}\delta_{im}\delta_{\alpha'_i\gamma_m}
 \expect{X_j^{\xi_j\eta'_j}}\delta_{jn}\delta_{\xi'_j\eta_n}
+\expect{X_i^{\alpha_i\eta'_n}}\delta_{in}\delta_{\alpha'_i\eta_n}
 \expect{X_j^{\xi_j\gamma'_m}}\delta_{jm}\delta_{\xi'_j\gamma_m} \nonumber \\
&&\hspace{-3cm}=\frac{\expo{-\beta \varepsilon_{\alpha_i}}}{z_i}
   \frac{\expo{-\beta \varepsilon_{\xi_j}}}{z_j}
   \left (-\delta_{im}\delta_{jn}
        \delta_{\alpha_i\gamma'_m}\delta_{\alpha'_i\gamma_m}
        \delta_{\xi_j\eta'_n}\delta_{\xi'_j\eta_n}
       +\delta_{in}\delta_{jm}
        \delta_{\alpha_i\eta'_n}\delta_{\alpha'_i\eta_n}
        \delta_{\xi_j\gamma'_m}\delta_{\xi'_j\gamma_m} \right ) \nonumber  
\eeq
Insertion into eq (\ref{Z2Hubbard}) shows, that only four out of the sixteen terms
survive. What we get finally is
\beq 
Z_2/Z_0&=& t^2 \summe{i\neq j}\summe{\sigma} 
\left (
 \frac{\expo{-\beta(\varepsilon_{i\sigma}+\varepsilon_{j0})}}{z_iz_j} 
       f_2\left( \varepsilon_{i\sigma} -\varepsilon_{i0}
                +\varepsilon_{j0} -\varepsilon_{j\sigma} \right)
\right. \nonumber\\
&&\hspace{0.7cm}
+\frac{\expo{-\beta(\varepsilon_{i\sigma}+\varepsilon_{j-\sigma})}}{z_iz_j} 
       f_2\left( \varepsilon_{i\sigma} -\varepsilon_{i0}
                +\varepsilon_{j-\sigma} -\varepsilon_{j2} \right)\nonumber\\
&&\hspace{0.7cm}
+\frac{\expo{-\beta(\varepsilon_{i2}+\varepsilon_{j0})}}{z_iz_j} 
       f_2\left( \varepsilon_{i2} -\varepsilon_{i-\sigma}
                +\varepsilon_{j0} -\varepsilon_{j\sigma} \right) \nonumber\\
&&\hspace{0.7cm}\left.
+\frac{\expo{-\beta(\varepsilon_{i2}+\varepsilon_{j-\sigma})}}{z_iz_j} 
       f_2\left( \varepsilon_{i2} -\varepsilon_{i-\sigma}
                +\varepsilon_{j-\sigma} -\varepsilon_{j2} \right) \right )
\eeq
With eqs (\ref{epsilonisigma}) we get therefore 
\beq 
Z_2/Z_0&=& t^2 \summe{i\neq j}\summe{\sigma} 
\left (
\frac{\expo{\beta(\mu+\sigma h_i)}}{z_iz_j}
    f_2\left(-\sigma (h_i-h_j)\right)
+\frac{\expo{\beta(2\mu+\sigma (h_i-h_j)}}{z_iz_j}
    f_2\left(-\sigma (h_i-h_j)-U\right)\right.\nonumber\\
&&\hspace{0.0cm}\left.
+ \frac{\expo{\beta(2\mu-U)}}{z_iz_j}
    f_2\left(U-\sigma (h_i-h_j)\right)
+ \frac{\expo{\beta(3\mu-U-\sigma h_j )}}{z_iz_j}
    f_2\left(-\sigma(h_i-h_j)\right)\right)
\label{Z2final}
\eeq
Although the above formula holds for arbitrary magnetic fields, we restrict
here to the most discussed ferromagnetic and antiferromagnetic cases and admit
nearest neighbour hopping only. 
For a homogenous magnetic field, i.e. $h_i=h_j=h$, we find
\beq
Z_2&=&Z_0 N g t^2 \summe{\sigma}
\left( 
\frac{\expo{\beta(\mu+\sigma h)}}{z^2(h)}f_2(0)+
\frac{\expo{\beta2\mu}}{z^2(h)}f_2(-U)
\right.\nonumber\\
&&\hspace{2.3cm}+\left.
\frac{\expo{\beta(2\mu-U)}}{z^2(h)}f_2(U)+
\frac{\expo{\beta(3\mu-U-\sigma h)}}{z^2(h)}f_2(0)
\right)
\eeq
Here N is the number of lattice sites and g the number of nearest neighbours.
For a staggered magnetic field, i.e. $h_i=-h_j$ with $h_i=h_s$ on the A
sublattice and $h_i=-h_s$ on the B sublattice, we have
\beq
Z_2&=&Z_0 N g t^2 \summe{\lambda}
\left( 
\frac{\expo{\beta(\mu+\lambda h_s)}}{z^2 (h_s)}f_2(-2\lambda h_s)+
\frac{\expo{\beta(2\mu+2\lambda h_s)}}{z^2(h_s)}f_2(-2\lambda h_s-U)
\right.\nonumber\\
&&
\hspace{2.3cm}+\left. 
\frac{\expo{\beta(2\mu-U)}}{z^2(h_s)}f_2(U-2\lambda h_s)+
\frac{\expo{\beta(3\mu-U-\lambda h_s)}}{z^2(h_s)}f_2(-2\lambda h_s)
\right)
\eeq 
with 
\beq
\lambda=\left\{\begin{array}{r@{\qquad}l}
1&\mbox{for $i\in$ A and }\sigma=+1\quad\mbox{or for $i\in$ B and }\sigma=-1\\
-1&\mbox{for $i\in$ A and }\sigma=-1\quad\mbox{or for $i\in$ B and }\sigma=+1
\end{array}\right.
\eeq
We compared our second order result given in eq (\ref{Z2final}) to 
that given in \cite{Plischke74,Metzner91,Pan91,Pan97} and found it in complete
accordance. Here we restrict ourself to the second order since we believe
it is enough to demonstrate the reliability of the presented perturbation 
theory. 
The calculation of higher order terms and discussion of the 
physics contained therein  we shift to a forthcoming paper. 
\section{Conclusion}
The main result of this paper is condensed in eq (\ref{fm}), since  it
contains the time integrations. We showed here the derivation via recursive
relations, since this way it is palpably that the degenerate case is included
and does not generate any problems. 
Furthermore, the recursive relation for the operators $\operator{a}_{nm}$ 
can be solved easily in symbolic manner by help of symbolic computer algebra
programs.
Once knowing the factor $f_m$, the remaining task, 
i.e. the calculation of the diagonal matrix 
elements, is straight forward. Since linked graphs have to be 
evaluated only, the further calculation steps are very likely to the coupled
cluster expansion. In \cite{Pan97} the authors did the time integrals 
by symbolic computation. In our theory this step is economized. At the first
glance one might think that this benefit is paid by the disadvantage that
one cannot permute the indices due to the weigth factors 
$f_m(\lambda_1,\dots,\lambda_{m-1})$. This is not the case 
due to the inherent symmetries of the functions $f_m$. Furthermore,
since these functions remain finite for arbitrary sets of energies, the
case of degeneration is included what is important for the typical models
of strong electron correlation. Nevertheless, 
it remains 
cumbersome enough to calculate all the linked diagrams in higher orders.
The reduction via commutations as shown in the appendix C is suited for
symbolic computer algebra. 
In case of the Hubbard model the result for every cluster is a sum of 
products of Kronecker symbols and 
single-site expectation values of the X-operators,
selecting a set of energy eigenvalues which specify the values of the
$\lambda's$ in the functions $f_m$. 
The form of the theory given in section 2 seems to exhibit slight
differences compared to the variant given in
section 4 for the Hubbard model. This stems from the fact, that in section
2 the perturbation was treated without knowledge of its inner structure,
whereas in section 4 we made use of this knowledge. If the special form
is known, one can calculate $V_{\mu\nu}$ as was demonstrated in appendix B,
thus showing both variants to be identical. 
\subsubsection*{Acknowledgement}
I am very grateful to Walter John and Roland Hayn for valueable discussion. 
\section*{Appendices}
\subsection*{Appendix A: Green function technique for eigenoperators}
\newcommand{\greenf}[2]{\langle\langle #1\,;\,#2\rangle\rangle}
\newcommand{\greenfo}[2]{\langle\langle #1\,;\,#2\rangle\rangle_{\omega}}
We define the retarded Greens function with respect to 
the operator $\hamilton_0$ in the usual way \cite{ElkGasser}
\beq
\greenf{\opA(t)}{\opB}&=&-i\Theta(t)\langle [\opA(t),\opB]_{+}\rangle
\eeq
$\Theta(t)$ is
the step function, and 
\beq
\opA(t)&=&\expo{it\hamilton_0}\opA \expo{-it\hamilton_0}=
\expo{it\liouville_0} \opA
\eeq
The Fourier-transformed equation of motion is than
\beq
\omega \greenfo{\opA}{\opB}&=&\langle [\opA,\opB]_{+}\rangle
-\greenfo{\liouville_0\opA}{\opB}
\eeq
Due to the eigenvalue equation holding for the eigenoperator $\opA$ this can
be solved for the wanted Greens function
\beq
\greenfo{\opA}{\opB}&=&\frac{\langle [\opA,\opB]_{+}\rangle}{\omega+\lambda_A}
\eeq
The expectation values $\langle\opB\opA\rangle$ one gets via the spectral 
theorem
\beq
\langle\opB\opA\rangle&=&
\lim_{\delta\rightarrow\infty} \frac{i}{2\pi}
\int_{-\infty}^{\infty}\frac{d \omega}{\expo{\beta\omega}+1}
\left \{ \greenf{\opA}{\opB}_{\omega+i\delta}
-\greenf{\opA}{\opB}_{\omega-i\delta}\right\} \nonumber \\
&=&
\frac{\langle [\opA,\opB]_{+}\rangle}{\expo{-\beta\lambda_A}+1}
\eeq
This is the same result as given in eq (\ref{anticommutatorfall}).
\subsection*{Appendix B: The straight way} 
First we calculate the matrix elements of $\operator{a}_{11}$.
To this end we define an eigenstate of $\hamilton_0$ by help of the X operators
in the following way
\beq
\mket{\{\mu\}}&=&\prod\limits_{l=1}^{N} X_l^{\mu_l0}\ketvacuum
\eeq
Here $\{\mu\}$ stands for a set $\{\mu_1,...,\mu_N\}$, 
summation over $\{\mu\}$
means N sums over the $\mu_i$.
The matrix elements to calculate are of the form
\beq
\bra{\{\mu\}}X_i^{\alpha_i\alpha'_i}X_j^{\xi_j\xi'_j}\ket{\{\nu\}}
\label{XX}
\eeq
Insertion of a unit operator yields
\beq
\summe{\{\gamma\}}\bra{\{\mu\}}X_i^{\alpha_i\alpha'_i}
\ket{\{\gamma\}}\bra{\{\gamma\}}X_j^{\xi_j\xi'_j}\ket{\{\nu\}}
\eeq
For the matrix element of a X-operator we find
\beq
\bra{\{\mu\}}X_i^{\alpha_i\alpha'_i}\ket{\{\nu\}}&=&\nonumber\\
&&\hspace{-2.5cm}
\bravacuum
\left(\prod\limits_{l=1}^{i-1} X_l^{0\mu_l}\right)
X_i^{0\nu_i}
\left(\prod\limits_{l=i+1}^{N} X_l^{0\mu_l}\right)
X_i^{\alpha_i\alpha'_i}
\left(\prod\limits_{l'=N}^{i+1} X_{l'}^{\nu_{l'}}\right)
X_i^{\nu_i0}
\left(\prod\limits_{l=i-1}^{1} X_{l'}^{\nu_{l'}}\right)
\ketvacuum
\eeq
Now commuting the $X_i$ operators to the center
yields
\beq
\bra{\{\mu\}}X_i^{\alpha_i\alpha'_i}\ket{\{\nu\}}&=&\nonumber\\
&&\hspace{-2.5cm}
(-1)^{P(\{\mu\}_{i+1}^N|0\mu_i)}(-1)^{P(\nu_i0|\{\nu\}_{N}^{i+1})}
\bravacuum\left(\prod\limits_{l\neq i}X_l^{0\mu_l}\right)
X_i^{0\mu_i}X_i^{\alpha_i\alpha'_i}X_i^{\nu_i0}
\left(\prod\limits_{l'\neq i}X_{l'}^{\nu_{l'}0}\right)\ketvacuum
\eeq
Here we used the parity function introduced in eq (\ref{unitedfall}).
The three X-operators at lattice site i implode according to
\beq  
X_i^{0\mu_i}X_i^{\alpha_i\alpha'_i}X_i^{\nu_i0}
&=&
X_i^{00}\delta_{\mu_i\alpha_i}\delta_{\alpha'_i\nu_i}
\eeq
Since $X_i^{00}$ is bosonic, moving it to the upmost right (or left)
will not change the sign. Furthermore it does not alter the vacuum, 
so that it may be omitted. What results is
\beq
\bra{\{\mu\}}X_i^{\alpha_i\alpha'_i}\ket{\{\nu\}}&=&\nonumber\\
&&\hspace{-2cm}
(-1)^{P(\{\mu\}_{i+1}^N|0\mu_i)}(-1)^{P(\nu_i0|\{\nu\}_{N}^{i+1})}
\delta_{\mu_i\alpha_i}\delta_{\alpha'_i\nu_i}
\bravacuum\left(\prod\limits_{l\neq i}X_l^{0\mu_l}\right)
\left(\prod\limits_{l'\neq i}X_{l'}^{\nu_{l'}0}\right)
\ketvacuum
\eeq
Due to the orthogonality of the eigenfunctions we find finally
\beq
\bra{\{\mu\}}X_i^{\alpha_i\alpha'_i}\ket{\{\nu\}}&=&
(-1)^{P(\{\mu\}_{i+1}^N|0\mu_i)}(-1)^{P(\nu_i0|\{\nu\}_{N}^{i+1})}
\delta_{\mu_i\alpha_i}\delta_{\alpha'_i\nu_i}
\prod\limits_{l\neq i}\delta_{\mu_l,\nu_l}
\eeq
Another way to simplify the same matrix element is to change 
$X_i^{\alpha_i\alpha'_i}$ either to the left till it is the right 
neighbour of $X_i^{0\mu_i}$ yielding
\beq
\bra{\{\mu\}}X_i^{\alpha_i\alpha'_i}\ket{\{\nu\}}&=&
(-1)^{P(\{\mu\}_{i+1}^N|\alpha_i\alpha'_i)}
\delta_{\mu_i\alpha_i}\delta_{\alpha'_i\nu_i}
\prod\limits_{l\neq i}\delta_{\mu_l,\nu_l}
\eeq
or to the right till it is the left neighbour of $X_i^{\nu_i0}$
resulting in
\beq
\bra{\{\mu\}}X_i^{\alpha_i\alpha'_i}\ket{\{\nu\}}&=&
(-1)^{P(\alpha_i\alpha'_i|\{\nu\}_{N}^{i+1})}
\delta_{\mu_i\alpha_i}\delta_{\alpha'_i\nu_i}
\prod\limits_{l\neq i}\delta_{\mu_l,\nu_l}
\eeq
The different forms can be created also by applying the following 
rules holding for the parity function $P(\opA|\opB)$
\beq
P(\opA|\opB)&=&P(\opB|\opA)\label{prule1}\\
P(\opA|\opB\opC)&=&P(\opA|\opC\opB)\label{prule2}\\
P(\opA|\opB\opC)&=&P(\opA|\opB)+P(\opA|\opC)\label{prule3}\\
P(\opA|\opB\opC)&=&P(\opA|\opB\opD\opD\opC)=
P(\opA|\opB\opD)+P(\opA|\opD\opC) \label{prule4}
\eeq
Thus we get for the matrix element in eq (\ref{XX})
\beq
\bra{\{\mu\}}X_i^{\alpha_i\alpha'_i}X_j^{\xi_j\xi'_j}\ket{\{\nu\}}
&\hspace{1em}&\nonumber\\
&&\hspace{-3cm} =\quad
\summe{\{\gamma\}}
(-1)^{P(\alpha_i\alpha'_i|\{\mu\}^{N}_{i+1})}
\delta_{\mu_i\alpha_i}\delta_{\alpha'_i\gamma_i}
\prod\limits_{l\neq i}\delta_{\mu_l,\gamma_l}
(-1)^{P(\xi_j\xi'_j|\{\nu\}_{N}^{j+1})}
\delta_{\gamma_j\xi_j}\delta_{\xi'_j\nu_j}
\prod\limits_{l'\neq j}\delta_{\gamma_{l'},\nu_{l'}}\nonumber\\
&&\hspace{-3cm}=\quad
(-1)^{P(\alpha_i\alpha'_i|\{\mu\}^{N}_{i+1})}
(-1)^{P(\xi_j\xi'_j|\{\nu\}_{N}^{j+1})}
\delta_{\mu_i\alpha_i}\delta_{\alpha'_i\nu_i}
\delta_{\mu_j\xi_j}\delta_{\xi'_j\nu_j}
\prod\limits_{l\neq i,j}\delta_{\mu_l,\nu_l}
\label{XXmunu}
\eeq
with this result we find the matrix elements of the perturbation to be
\beq
\bra{\{\mu\}}\hamilton_1\ket{\{\nu\}}
&=&
-t \prod_{l\neq i,j}\delta_{\mu_l\nu_l}
\summe{i\neq j}\summe{\sigma} 
(-1)^{P(f|\{\mu\}^{N}_{i+1})}
(-1)^{P(f|\{\nu\}_{N}^{j+1})}\times\nonumber \\
&&\times
\bigl (        \delta_{\mu_i\sigma}\delta_{0\nu_i}
       +\sigma \delta_{\mu_i2}\delta_{-\sigma\nu_i} \bigr )
\bigl (        \delta_{\mu_j0}\delta_{\sigma\nu_j}
       +\sigma \delta_{\mu_j-\sigma}\delta_{2\nu_i} \bigr )
 \eeq 
Here we used the fact that both $X_i^{\sigma0}$ and $X_i^{2-\sigma}$ are
fermionic and we abbreviated both with $f$.
The second order contribution to the partition sum reads now 
\beq
Z_{2}&=&t^2\summe{i\neq j}\summe{\sigma}\summe{m\neq n}\summe{\sigma'}
\summe{\{ \mu \}}\summe{\{ \nu \}}\expo{-\beta E_{\{ \mu \}}}
f_2(E_{\{ \mu \}}-E_{\{ \nu \}}) 
\prod_{l\neq i,j}\delta_{\mu_l\nu_l}\prod_{l'\neq i,j}\delta_{\mu_{l'}\nu_{l'}}
\times \\
&&\hspace{0.5cm} \times
(-1)^{P(f|\{\mu\}^{N}_{i+1})}
(-1)^{P(f|\{\nu\}_{N}^{j+1})}
(-1)^{P(f|\{\nu\}^{N}_{m+1})}
(-1)^{P(f|\{\mu\}_{N}^{n+1})}
\nonumber \\
&&\hspace{0.5cm} \times
\bigl (        \delta_{\mu_i\sigma}\delta_{0\nu_i}
       +\sigma \delta_{\mu_i2}\delta_{-\sigma\nu_i} \bigr )
\bigl (        \delta_{\mu_j0}\delta_{\sigma\nu_j}
       +\sigma \delta_{\mu_j-\sigma}\delta_{2\nu_i} \bigr )
\nonumber \\
&&\hspace{0.5cm} \times
\bigl (        \delta_{\nu_m\sigma'}\delta_{0\mu_m}
       +\sigma' \delta_{\nu_m2}\delta_{-\sigma'\mu_m} \bigr )
\bigl (        \delta_{\nu_n0}\delta_{\sigma'\mu_n}
       +\sigma' \delta_{\nu_n-\sigma'}\delta_{2\mu_n'} \bigr )
\nonumber 
\eeq
Next we carry out the $\{\nu\}$-sum.
All contributions where the indices i,j are distinct from m and n
vanishes due to the fact that the X-operators within the
perturbation are nondiagonal.
What remains are the two contributions according 
to i=m, j=n and i=n, j=m. 
The first contribution also vanishes, what may be shown by doing
the same steps we shall employ in the following for the calculation of 
the nonvanishing second case. 
Introducing the shortcut $\{\mu\}_{/ij}$ for the set of $\mu$'s
except $\mu_i$ and $\mu_j$ we find
\beq
Z_2&=& t^2\summe{i\neq j}\summe{\sigma}
\summe{\{\mu\}_{/ij}}\expo{-\beta E(\{\mu\}_{/ij})}
\summe{\mu_i\mu_j}\summe{\nu_i\nu_j}
\expo{-\beta(\varepsilon_{\mu_i}+\varepsilon_{\mu_j})}
f_2(\varepsilon_{\mu_i}+\varepsilon_{\mu_j}-
    \varepsilon_{\nu_i}+\varepsilon_{\nu_j})
\times\nonumber \\
&&\hspace{0.0cm}\times
(-1)^{P(f|\{\mu\}^{N}_{i+1})}(-1)^{P(f|\{\mu\}_{N}^{i+1})}\nonumber \\
&&\hspace{0.0cm}\times
\left ( 
\delta_{j>i}(-1)^{P(f|\{\mu\}_{N}^{j+1})}(-1)^{P(f|\{\mu\}^{N}_{j+1})}+
\delta_{j<i}(-1)^{P(f|\{\mu\}_{N}^{i+1}\nu_i\{\mu\}_{i-1}^{j+1})}
            (-1)^{P(f|\{\mu\}_{N}^{i+1}\nu_i\{\mu\}_{i-1}^{j+1})}
\right )
\nonumber \\
&&\hspace{0.0cm} \times
\bigl (        \delta_{\mu_i\sigma}\delta_{0\nu_i}
       +\sigma \delta_{\mu_i2}\delta_{-\sigma\nu_i} \bigr )
\bigl (        \delta_{\mu_j0}\delta_{\sigma\nu_j}
       +\sigma \delta_{\mu_j-\sigma}\delta_{2\nu_i} \bigr )
\nonumber \\
&&\hspace{0.0cm} \times
\bigl (        \delta_{\nu_j\sigma'}\delta_{0\mu_j}
       +\sigma' \delta_{\nu_j2}\delta_{-\sigma'\mu_j} \bigr )
\bigl (        \delta_{\nu_i0}\delta_{\sigma'\mu_i}
       +\sigma' \delta_{\nu_i-\sigma'}\delta_{2\mu_i'} \bigr )
\nonumber 
\eeq
By help of the rules given for the parity function 
eqs (\ref{prule1},\ref{prule2},\ref{prule3},\ref{prule4}) it is easy to show
that we get a plus sign.
Expanding the products, we find that only four out of the sixteen terms
survive
\beq
Z_2&=& t^2 \summe{i\neq j}\summe{\sigma}
\summe{\{\mu\}_{/ij}}\expo{-\beta E(\{\mu\}_{/ij})}
\summe{\mu_i\mu_j}\summe{\nu_i\nu_j}
\expo{-\beta(\varepsilon_{\mu_i}+\varepsilon_{\mu_j})}
f_2(\varepsilon_{\mu_i}+\varepsilon_{\mu_j}-
    \varepsilon_{\nu_i}+\varepsilon_{\nu_j})
\times\nonumber \\
&&\hspace{0.0cm} \times
\bigl (
\delta_{\mu_i\sigma}\delta_{0\nu_i}\delta_{\sigma\sigma'}
+\sigma \sigma'\delta_{\mu_i2}\delta_{-\sigma\nu_i}\delta_{\sigma\sigma'} 
\bigr ) 
\bigl (        
\delta_{\mu_j0}\delta_{\sigma\nu_j}\delta_{\sigma\sigma'}
       +\sigma \delta_{\mu_j-\sigma}\delta_{2\nu_i} \bigr )
+\sigma \sigma'\delta_{\mu_j2}\delta_{-\sigma\nu_j}\delta_{\sigma\sigma'}
\bigr ) 
\eeq
Summing over $\mu_i,\mu_j,\nu_i,\nu_j $ and $\sigma'$ yields
\beq
Z_2&=t^2&\summe{\{\mu\}_{/ij}}\expo{-\beta E(\{\mu\}_{/ij})}
\summe{i\neq j}\summe{\sigma}\times\nonumber \\
&&\times
\left \{
\expo{-\beta(\varepsilon_{i\sigma}+\varepsilon_{j0})}
f_2(\varepsilon_{i\sigma}+\varepsilon_{j0}-
    \varepsilon_{i0}-\varepsilon_{j\sigma})
\right .\nonumber \\
&&\hspace{0.0cm}+
\expo{-\beta(\varepsilon_{i\sigma}+\varepsilon_{j-\sigma})}
f_2(\varepsilon_{i\sigma}+\varepsilon_{j-\sigma}-
    \varepsilon_{i0}-\varepsilon_{j2})\nonumber \\
&&\hspace{0.0cm}+
\expo{-\beta(\varepsilon_{i2}+\varepsilon_{j0})}
f_2(\varepsilon_{i2}+\varepsilon_{j0}-
    \varepsilon_{i-\sigma}-\varepsilon_{j\sigma})\nonumber \\
&&\hspace{0.0cm}+
\left .
\expo{-\beta(\varepsilon_{i2}+\varepsilon_{j-\sigma})}
f_2(\varepsilon_{i2}+\varepsilon_{j-\sigma}-
    \varepsilon_{i-\sigma}-\varepsilon_{j2})
\right \}  
\eeq
The sum over $\{\mu\}_{/ij}$ may be expressed by the unperturbed
partition sum according to
\beq
\summe{\{\mu\}_{/ij}}\expo{-\beta E(\{\mu\}_{/ij})}
&=&\frac{Z_0}{z_iz_j}
\eeq
with $z_i$ beeing the unperturbed single site partition sum.
Thus we get exactly the same result of the second order
contribution to the partition sum as was given in eq (\ref{Z2final}).
The reader might have got the impression, that the method is 
complicated due to the multitude of factors $(-1)^P(\{\cdots\}|\{\cdots\})$ 
containing dummy
lattice sites. This is a result of our aim to demonstrate here
the straight forward character of the theory instead of modifying  it to the
special case of the Hubbard model. The straight calculation consists of
firstly writing down the expression of $Z_m$ in dependence of the 
$\lambda_1,\cdots,\lambda_{m-1}$ , secondly one has to calculate the 
matrix elements of the perturbation with respect to the unperturbed 
(many body) states, and thirdly one has to multiply the  m$^{th}$ power of
that matrix to the known function $f_m(\lambda_1,\cdots,\lambda_{m-1})$.
This way we have not to evaluate any graphs or difficult Green functions,
instead  simple matrix multiplications have to be carried out, thereby 
specifying the $\lambda$'s, what 
is very comfortable as long as the dimension of the matrix is not to large,
what is the case for instance in small cluster problems. In condensed matter
systems, especially if one is interested in the limit $N\rightarrow\infty$,
other methods can be utilized, as was shown for the Hubbard model in
section 4.
\subsection*{Appendix C: Systematic evaluation via commutations}
For an automated calculation Wick's theorem for the X-operators is not very 
convenient. A more
systematic calculation starts from eq (\ref{unitedfall}). 
Again it should be demonstrated for the evaluation of the second
order expectation values, having the form given in eq (\ref{XXXX}).
In the first step all X-operators are fermionic and we get therefore
\beq
\expect{X_i^{\alpha\alpha'}X_j^{\xi\xi'} 
        X_m^{\gamma\gamma'}X_n^{\eta\eta'}}
&=&
\frac{1}{\expo{\beta\lambda_i^{\alpha\alpha'}}+1}
\expect{[X_i^{\alpha\alpha'},X_j^{\xi\xi'} 
        X_m^{\gamma\gamma'}X_n^{\eta\eta'}]_{+}} \nonumber \\
&=&   
\frac{ -\delta_{im}}{\expo{\beta\lambda_i^{\alpha\alpha'}}+1}
\left ( \delta_{\alpha'\gamma}
        \expect{X_j^{\xi\xi'}X_i^{\alpha\gamma'}X_n^{\eta\eta'}}
       +\delta_{\alpha\gamma'}
        \expect{X_j^{\xi\xi'}X_i^{\gamma\alpha'}X_n^{\eta\eta'}}
\right ) \nonumber \\
&&+\frac{ \delta_{in}}{\expo{\beta\lambda_i^{\alpha\alpha'}}+1}
\left ( \delta_{\alpha'\eta}
        \expect{X_j^{\xi\xi'}X_m^{\gamma\gamma'}X_i^{\alpha\eta'}}
       +\delta_{\alpha\eta'}
        \expect{X_j^{\xi\xi'}X_m^{\gamma\gamma'}X_i^{\eta\alpha'}}
\right )
\eeq
Here we used the fact $i \neq j$.
In the next step we have to calculate the product of three X-operators.
Since in the general form used here, it is possible that e.g. 
$X_i^{\alpha\gamma'}$ is either fermionic or bosonic, we have to take into
account the parity function. As an example we show here the calculation
of only one of these expectation values 
\beq
\expect{X_j^{\xi\xi'}X_i^{\alpha\gamma'}X_n^{\eta\eta'}}
&=&
\frac{1}{1-g
\expo{\beta\lambda_j^{\xi\xi'}}}
\expect{[X_j^{\xi\xi'}, 
        X_i^{\alpha\gamma'}X_n^{\eta\eta'}]_{g}} \nonumber \\
&=&   
\frac{ (-1)^{P(X_j^{\xi\xi'}|X_i^{\alpha\gamma'})}\delta_{jn}}{1-g\expo{\beta\lambda_j^{\xi\xi'}}}
\left (        \delta_{\xi'\eta}
        \expect{X_i^{\alpha\gamma'}X_j^{\xi\eta'}}
       +\delta_{\xi\eta'}
        \expect{X_i^{\gamma\alpha'}X_j^{\eta\xi'} }
\right )
\\
g&=&(-1)^{P(X_j^{\xi\xi'}|X_i^{\alpha\gamma'}X_n^{\eta\eta'})} \nonumber 
\eeq
Since we have $i \neq j $
the operators $X_i^{\gamma\alpha'}$ and $X_j^{\eta\xi'}$ either 
commute or anticommute. It follows from the statements given in section 3
that the expectation value of two (anti-)commuting X-operators 
at different lattice sites 
$\expect{X_i^{\alpha\alpha'}X_j^{\xi\xi'}}$ is zero except 
that  $\lambda_i^{\alpha\alpha'}=0$ and 
$\lambda_j^{\xi\xi'}=0$ hold simultaneously, what means here that we have
$\varepsilon_{\alpha}=\varepsilon_{\alpha'}$ 
and $\varepsilon_{\xi}=\varepsilon_{\xi'}$
or equivalently $\alpha=\alpha'$ and $\xi=\xi'$. In a condensed form this reads
\beq 
\expect{X_i^{\alpha\alpha'}X_j^{\xi\xi'}}
&=&
\expect{X_i^{\alpha\alpha}X_j^{\xi\xi}} \delta_{\alpha\alpha'}\delta_{\xi\xi'}
\eeq
For the remaining expectation value $\expect{X_i^{\alpha\alpha}X_j^{\xi\xi}}$
one finds by a direct calculation
using eq (\ref{XXmunu}) 
$X_i^{\alpha\alpha}$ and $X_j^{\xi\xi}$
\beq
\expect{X_i^{\alpha\alpha}X_j^{\xi\xi}}
&=&
\frac{\expo{-\beta\varepsilon_{i\alpha}}}{z_i}
\frac{\expo{-\beta\varepsilon_{j\xi}}}{z_j}
\eeq
\subsection*{Appendix D: An alternate derivation of $f_m$}

This alternate derivation was given by Walter John after critical reading the
manuscript of this paper.\\

In the following we start from 
\beq
f_m(\beta;\lambda_1,\dots,\lambda_{m-1})\!\!&=&\!\!
\int\limits_{0}^{\beta}\!d\beta_1\int\limits_{0}^{\beta_1}\!d\beta_2 \cdots
\!\!\int\limits_{0}^{\beta_{m-1}}\!\!\!d\beta_m
\expo{\beta_1\lambda_1}
\expo{\beta_2(-\lambda_1+\lambda_2)}
\cdots
\expo{\beta_m(-\lambda_{m-1} )} \label{fmwithbeta}
\eeq
Here we mentioned the parameter $\beta$ explicitely. The Laplace transformed
function is then
\beq
F_m(p;\lambda_1,\dots,\lambda_{m-1})\!\!&=&\!\!
\int\limits_{0}^{\infty}\!d\beta \expo{-p\beta}
f_m(\beta;\lambda_1,\dots,\lambda_{m-1})
\eeq
Partial integration yields
\beq
F_m(p;\lambda_1,\dots,\lambda_{m-1}) 
&=& 
\left [ 
\frac{\expo{-p\beta}}{-p}f_m(\beta;\lambda_1,\dots,\lambda_{m-1})
\right ]_{0}^{\infty}  \\
&&\hspace{-2cm}+\,\frac{1}{p}
\int\limits_{0}^{\infty}\!d\beta \expo{-(p-\lambda_1)\beta}
\int\limits_{0}^{\beta}\!d\beta_2
\expo{\beta_2(-\lambda_1+\lambda_2)}
\int\limits_{0}^{\beta_2}\!d\beta_3 \cdots
\!\!\int\limits_{0}^{\beta_{m-1}}\!\!\!d\beta_m
\expo{\beta_m(-\lambda_{m-1} )}\nonumber  
\eeq
The first term on the right hand side vanishes, if $p$ is chosen adequately.
Integrating by parts another time delivers
\beq
F_m(p;\lambda_1,\dots,\lambda_{m-1}) 
&=& \\
&&\hspace{-2cm} 
\frac{1}{p}\,\frac{1}{p-\lambda_1}
\int\limits_{0}^{\infty}\!d\beta \expo{-(p-\lambda_2)\beta}
\int\limits_{0}^{\beta}\!d\beta_3
\expo{\beta_3(-\lambda_2+\lambda_3)}
\int\limits_{0}^{\beta_3}\!d\beta_4 \cdots
\!\!\int\limits_{0}^{\beta_{m-1}}\!\!\!d\beta_m
\expo{\beta_m(-\lambda_{m-1} )}\nonumber 
\eeq
Thus from repeated partial integrations one gets finally
\beq
F_m(p;\lambda_1,\dots,\lambda_{m-1}) 
&=& 
\frac{1}{p^2}\prod_{i=1}^{m-1}\frac{1}{p-\lambda_i}
\eeq
The original function one gets via back-transforming
\beq
f_m(\beta;\lambda_1,\dots,\lambda_{m-1}) &=& \frac{1}{2\pi i}
\int\limits_{C-i\infty}^{C+i\infty}\!dp \, \expo{p\beta}
F_m(p;\lambda_1,\dots,\lambda_{m-1})
\eeq
Here C is a real constant larger than the maximum $\lambda_i$. 
The integration path may be deformed to encircle the individual poles
lying on the real axes. Thus we get from Cauchy's theorem
\beq
f_m(\beta;\lambda_1,\dots,\lambda_{m-1})&=&
\sum_{i=1}^{m-1}\frac{\expo{\beta\lambda_i}}{\lambda_i^2}
\frac{1}{\prod_{j\neq i}^{m-1}(\lambda_i-\lambda_j)}
+
\frac{1}{\prod_{j=1}^{m-1}(-\lambda_j)}
\left (\beta + \sum_{i=1}^{m-1}\frac{1}{\lambda_i}\right )
\eeq
The second term on the right hand side stems from the double pole at zero.
Here we admitted only simple poles to get a concise form. There is no
problem with the degenerate case, since in case that a pole has a higher
order, one has to take higher derivatives during application of
Cauchy's theorem. Via substitution of the exponential factor by help of eq 
(\ref{Em}) one can show the derived formula to be equivalent to eq
(\ref{fm}).


\begin{thebibliography}{99}
\bibitem{Hubbard63} J. Hubbard, Proc.R.Soc.London, Ser. {\bf A 236} (1963) 238
\bibitem{Gutzwiller63} M.C. Gutzwiller, Phys.Rev.Lett. {\bf 10} (1963) 159
\bibitem{FuldeBuch} P. Fulde. Electron Correlations in Molecules and Solids.
\bibitem{Kubo80} K. Kubo, Prog.Theor.Phys. {\bf 64} (1980) 758
Springer Verlag, third edition, 1995
\bibitem{RKubo62}R. Kubo, J. Phys. Soc. Jpn {\bf 17} (1962) 1100
\bibitem{Becker87} K. W. Becker and P. Fulde, Z. Phys.{\bf B 72} (1988) 423;
see also  \cite{FuldeBuch} 
\bibitem{Kladko97}K. Kladko and P. Fulde, cond-mat/9709044 
\bibitem{Espinosa96}A. W. Espinosa Mueller and A. R. Matamala Vasquez,
         quant-ph/9606011 and quant-ph/9606013, Version Nov. 6, 1997
\bibitem{Schork92} T. Schork and P. Fulde, J. Chem. Phys. {\bf 97} (1992)9195 
\bibitem{CCmethod} in nuclear physics: F. Coester and H. Kuemmel, Nucl. Phys.
         {\bf 17} (1960), 
	in quantum chemistry: J. \v{C}i\v{z}ek, 
        J. Chem. Phys. {\bf 45} (1966) 4256
\bibitem{ElkGasser}K. Elk, W. Gasser, Die Methode der Greenschen Funktionen
	in der Festkoerperphysik, Akademie Verlag, Berlin, 1979
\bibitem{Beni73} G. Beni, P. Pincus, and D. Hone, 
                 Phys.Rev. B {\bf 8} (1983) 3389
\bibitem{Plischke74} M. Plischke, J.Stat.Phys {\bf 11} (1974) 159
\bibitem{Zhao87} B.H. Zhao, H.Q. Nie, K.Y. Zhang, K.A. Chao, and R. Minas, 
                 Phys.Rev. B {\bf36} (1983) 2231
\bibitem{Metzner91} W. Metzner, Phys.Rev. B {\bf 43} (1991) 8549 
\bibitem{Kubo83} K. Kubo and M. Tada, Prog.Theor.Physics {\bf 69} (1983)1345
\bibitem{Kubo84} K. Kubo and M. Tada, Prog.Theor.Physics {\bf 71} (1984)479
\bibitem{Pan91} K.K. Pan and Y.L. Wang, Phys.Rev. B {\bf 43} (1991)3706
\bibitem{Pan97} K.K. Pan and Y.L. Wang, Phys.Rev. B {\bf 55} (1997)2981
\bibitem{John97} W. John, private communication
\bibitem{Abrikosov}A.A. Abrikosov, L.P. Gorkov, and I.Ye. Dzyaloshinsky,
        {\it Quantum Field Theoretical Methods in 
                                     Statistical Physics} (Pergamon New York) 
\bibitem{Hubbard65} J. Hubbard, Proc.R.Soc.London, Ser. {\bf A 285} (1965) 542
\bibitem{Isjumov87} Ju.A. Isjumov and Ju.N. Skrjabin, 
                Statistical Mechanics of ordered magnetic systems (in Russian), 
                Nauka , Moscow  1997
\end{thebibliography}
\end{document}